\documentclass[]{raa}            % referee version: for submission
\usepackage{graphicx,times}
\usepackage{natbib}

\providecommand{\bm}[1]{\mbox{\boldmath$#1$\unboldmath}}

\begin{document}

\title{Spectroscopic Study of Globular Clusters in the Halo of
  M31 with Xinglong 2.16m Telescope
% $^*$
%\footnotetext{\small $*$ Supported by the National Natural Science Foundation of China.}
}

\volnopage{ {\bf 2011} Vol.\ {\bf 9} No. {\bf 00}, 000--000}
\setcounter{page}{1}

\author{Zhou Fan
  \inst{1,2}
  \and Ya-Fang Huang
  \inst{1,2}
  \and Jin-Zeng Li
  \inst{1,2}
  \and Xu Zhou
  \inst{1,2}
  \and Jun Ma
  \inst{1,2}
  \and Hong Wu
  \inst{1,2}
  \and Tian-Meng Zhang
  \inst{1,2}
  \and Yong-Heng Zhao
  \inst{1,2}
}

\institute{National Astronomical Observatories, Chinese Academy of Sciences,
  Beijing 100012, China; {\it zfan@bao.ac.cn}\\
  \and
  Key Laboratory of Optical Astronomy, National Astronomical
  Observatories, Chinese Academy of Sciences
  \vs \no
  {\small Received [year] [month] [day]; accepted [year] [month] [day] }
}

\abstract{We present the spectroscopic observations for 11
confirmed globular clusters of M31 with the OMR spectrograph on 2.16m
telescope at Xinglong site of National Astronomical Observatories,
Chinese Academy of Sciences. Nine of our sample clusters are located
in the halo of M31 and the most remote one is out to a projected
radius of 78.75 kpc from the galactic center. For all our sample
clusters, we measured the  Lick absorption-line indices and the radial
velocities. It is noted that most GCs of our sample are distinct from
the HI rotation curve of M31 galaxy, especially for B514, MCGC5, H12
and B517, suggesting that most of our sample clusters do not have
kinematic association with the star forming young disk of the galaxy. 
We fitted the absorption line indices with the updated stellar
population model \citet{tmj} with two different tracks of Cassisi and
Padova, separately, by applying the $\chi^2-$minimization method. The
fitting results show that all our sample clusters are older than 10
Gyr, and metal-poor ($-0.91 \le $ [Fe/H] $\le -2.38$ dex).  
After merging the spectroscopic metallicity of our work with the
previously published ones, we extended the cluster sample out to a
projected radius of 117 kpc from the galaxy's center. We found the 
metallicity gradient for all the confirmed clusters exists with a
slope of $-0.028\pm0.001$ dex kpc$^{-1}$. However, the slope turns to
be $-0.018\pm0.001$ dex kpc$^{-1}$ for all the halo clusters, which is
much shallower. If we only consider the outer halo clusters with
$r_{\rm p}>25$ kpc, the slope becomes $-0.010\pm0.002$ dex kpc$^{-1}$
and if one cluster G001 is excluded from the outer halo sample,  the
slope is $-0.004\pm0.002$ dex kpc$^{-1}$. Thus we conclude
that metallicity gradient for M31 outer halo clusters is not
significant, which agrees well with the previous findings.  } 

   \authorrunning{Fan et al.}            %author_head in even pages
   \titlerunning{Spectroscopic Study of M31 Halo GCs}  % title_head in odd pages
   \maketitle

%% The author head (on even pages) and the title head (on odd pages) will be
%% automatically extracted from \author{} and \title{}. Whenever the title is too long,
%% you will be asked to supply a shorter one by inserting either \authorrunning{} or
%% \titlerunning{} before \maketitle. Anyway, you can specify your own heads.
%%
%%
%% Note: In the following text body of your manuscript, please note several differences from
%%       other major journals:
%% (1) \subsection{Please Capitalize the First Letter of Each Notional Word in Subsection Title}
%% (2) Please Capitalize the First Letter of Each Notional Word in all tables' captions

%
%________________________________________________ sections below
%
\section{Introduction}           %% first-level sections will be auto-capitalized
\label{intro.sec}

Galactic formation and evolutionary scenarios remain among the most
important outstanding problems in contemporary astrophysics
\citep{per02}. One way to better understand these questions is through
detailed studies of globular clusters (GCs). These objects are often
considered fossils of galactic formation and evolution processes,
since they formed at the very early stages of their host galaxies'
lifecycles \citep{bh00}. GCs are usually densely packed,
gravitationally bound spherical systems containing several thousands
to approximately one million stars. Thus, they can be detected from
great distances and are suitable as probes for studying the properties
of extragalactic systems.

Located at a distance of approximately 780 kpc \citep{sg98, mac01,mcc05},
M31 is the nearest and largest spiral galaxy in our Local Group. It
contains a large number of GCs and is considered an ideal laboratory
for studies of star clusters in external galaxies. \citet{bh01}
estimated the total number of GCs at 
$460\pm70$, while \citet{per10} arrived at $\sim$530. Both of these
estimates yield much larger numbers than for the GCs in our
Galaxy. However, from the observational evidence collected to date
\citep[see, e.g.,][]{rich05}, the M31 GCs and their Galactic
counterparts reveal some striking similarities
\citep{ffp94,dj97,bhh02}. Based on survey data from the
Canada-France-Hawaii Telescope (CFHT) and the Wide Field Camera on the
Isaac Newton Telescope (INT), \citet{h07} concluded that M31 and the
Milky Way are more similar than previously thought. However, \citet{ham07}
compared our Galaxy and M31 to the local disk galaxies within the
same mass range and found that Milky Way is an exceptional disk galaxy
which did not undergo any significant merging for last 10 Gyrs so that it
lacks stellar mass, angular momentum, disk radius ad metallicity of
stars in the outskirts while M31 is a typical disk galaxy which is
shaped by relatively recent merging. This may explain why there are
more GCs in M31 than that in our Galaxy by a factor of 3 as the
merging could lead to the formations of GCs. Later, \citet{yin09} found that 
the two galaxies are similar in the radial profiles of star formation
rate, gas profiles and stellar metallicity distributions along the
disk by studying the chemical evolution history of the two galaxies.
The authors concluded that the star formation efficiency of M31 disk
is twice high as that in our Galaxy. \citet{hou09} also compared the
two disk galaxies and concluded that the Milky Way disk contains more
gas and higher star formation rate than that of M31.  The authors also
find that the scaled abundance gradients are similar for the two
galaxies. These recent works could provide useful clues which can
explain the similarities and differences of two GC system. Therefore,
studying the properties of the GCs in M31 not 
only improves our understanding of the formation and structure of our
nearest large neighbor, but also of our own Galaxy. 

A large number of halo GCs in M31 have recently been discovered. These
are important to study the formation history of M31 and its dark
matter content. \citet{h04} discovered nine previously unknown GCs in
the halo of M31 using the INT survey. Subsequently, \citet{h05} found
three new, extended GCs in the halo of M31, which have characteristics
between typical GCs and dwarf galaxies. \citet{mac06} reported four
extended, low-surface-brightness clusters in the halo of M31 based on
{\sl Hubble Space Telescope}/Advanced Camera for Surveys (ACS)
imaging. They are structurally very different from typical M31
GCs. However, their old and metal-poor characteristics are similar to
those of typical GCs. \citet{h07} discovered 40 new GCs in the halo of
M31 (out to 100 kpc from the galactic center) based on INT and CFHT
imaging. Some of them are also very extended. These extended
star clusters in the M31 halo are very similar to the diffuse star
clusters (DSCs) associated with early-type galaxies in the Virgo
Cluster reported by \citet{peng06} based on the ACS Virgo Cluster
survey. However, it seems that DSCs are usually fainter than typical
GCs. \citet{mac07} reported 10 outer-halo GCs in M31, at $\sim$15 kpc
to 100 kpc from the galactic center. Eight of these were newly
discovered based on deep ACS imaging. The halo GCs in their sample are
very bright, compact, and metal poor, and therefore quite different
from their counterparts in our Galaxy. \citet{ma10}  constrained the
age, metallicity, reddening and distance modulus of B379, which is
located in the halo of M31, with the SSP model and photometry.

In this paper, we will present our new observations on a sample
  of new GCs, most of them are located far from the galaxy
  center. This allows us to be able to study the properties of the M31
  outer halo in more detail. The paper 
is organized as follows. In \S \ref{sam.sec} we describe how 
we selected our sample of M31 GCs and their spatial distribution. In
\S \ref{obs.sec}, we reported the spectroscopic observations with 2.16
m telescope at Xinglong site and the data reductions from which the
radial velocities and Lick line indices were measured.
  Subsequently, in \S \ref{fit.sec}, we derive the ages and
  metallicities of 
  GCs with $\chi^2-$minimization fitting. We also discuss our final
  results on the metallicity distribution in the M31 halo. We give our
  summary in \S \ref{sum.sec}.

\section{Sample selection}
\label{sam.sec}

We selected the sources from the Revised Bologna Catalogue of M31  
globular clusters and candidates \citep[RBC
v.4, available from http://www.bo.astro.it/M31;][]
{gall04,gall06,gall07,gall09} , which is the latest and most
comprehensive M31 GC catalogue so far. It contains 2045 objects, 
including 663 confirmed star clusters, 604 cluster candidates, and 
778 other objects that were initially thought to be GCs but later 
proved to be stars, asterisms, galaxies, or H{\sc ii} regions. In fact,
many of the halo clusters were from \citet{mac07}, who reported 10 GCs in 
the outer halo of M31 from their deep ACS images, of which eight were 
detected for the first time (see for details in \S \ref{intro.sec}). 
In our work, our sample clusters are completely selected from RBC
v.4. We selected the confirmed and bright ($<17$ mag in $V$ band)
clusters as well as being located as far as they could from the galaxy
center, where the local background is too bright to observe. Finally,
we have 11 bright confirmed clusters in our sample and most of them
are located in the halo of the galaxy.  
Although some of our sample clusters have the previous spectroscopic
observations by some authors, actually those clusters lack
comprehensive spectroscopic informations. In other words, they
only have radial velocities or [Fe/H] or [$\alpha$/Fe] or age
informations. Therefore it is necessary to observe the spectra of our
sample clusters systematically and study the ages and metallicity in detail. 

The informations of our sample GCs are listed in Table~\ref{t1.tab}, including
coordinates, projected radii, $V$-band magnitudes and age estimates. 
All the coordinates (Cols. 2 and 3) and $V$-band magnitudes (Col. 5)
are from RBC v.4 except the $V$ mag of EXT8 which was derived from
$ugriz$ photometry of \citet{pc10} with the transformation equation of
\citet{jes05} as $V$ mag of EXT8 is not provided in RBC v.4. The
projected radii from the galaxy center $r_{\rm p}$ (Col. 4) were
calculated with M31 center coordinate $00:42:44.31, +41:16:09.4$
\citep{per02}, PA $=38^{\circ}$ and distance $=785$ kpc
\citep{mcc05}. The ages (Col. 6) are from a number of previous work:
\citet{pc10} by using the SDSS and 2MASS photometric colors;
\citet{cw09} by using the 6.5m MMT Hectospec spetra line indices and HST
CMD fittings, \citet{gall05} by comparing the lines indices with the
prediction models. 

\begin{table}[ht!!!]
\small
\centering
\begin{minipage}[]{100mm}
\caption[]{The parameters of our sample GCs.}\label{t1.tab}\end{minipage}
%%Please Capitalize the First Letter of Each Notional Word in table's caption
\tabcolsep 3mm
\begin{tabular}{lcccccc}
  \hline\noalign{\smallskip}
  ID & R.A. & Dec. &$r_{\rm p}$&$V$ & age & references for ages$^b$\\
  & (J2000) & (J2000) &(kpc)&(mag) & & \\
  \hline\noalign{\smallskip}
  MCGC2   & 00:29:44.90  & +41:13:09.8 & 33.47& 16.98& old& P     \\
  MCGC3   & 00:30:27.30  & +41:36:20.4 & 31.88& 16.31& old& P    \\
  B514       & 00:31:09.90  & +37:53:59.7 & 55.39& 15.76& $>10$ Gyr& G\\
  MCGC5   & 00:35:59.73  & +35:41:03.8 & 78.73& 16.09& old& P     \\
  B298       & 00:38:00.23  & +40:43:55.9 & 14.28& 16.59& old& C\\
  H12        & 00:38:03.85  & +37:44:00.6 & 50.03& 16.47&   & \\
  B019       & 00:40:52.52  & +41:18:53.4 &  4.84& 14.93& old& C\\
  B020       & 00:40:55.26  & +41:41:25.2 &  7.42& 14.91& interm /
  old& P, C\\
  B023       & 00:41:01.18  & +41:13:45.7 &  4.46& 14.22& old&  P, C \\
  EXT8       & 00:53:14.51  & +41:33:24.7 & 27.27& 15.54$^a$   & & \\
  B517       & 00:59:59.91  & +41:54:06.6 & 45.08& 16.08&  & \\
  \noalign{\smallskip}\hline
\end{tabular}
\tablecomments{0.86\textwidth}{ $r_{\rm p}$ refers to the projected
  radius from the center of the galaxy.\\
  $^a$ derived from $ugriz$ photometry of \citet{pc10} with the
  transformation equation of \citet{jes05}.\\
  $^b$ P: age estimates from \citet{pc10};  C: from \citet{cw09}; G: from \citet{gall05}. }
\end{table}

\begin{figure}
\resizebox{\hsize}{!}{\rotatebox{0}{\includegraphics{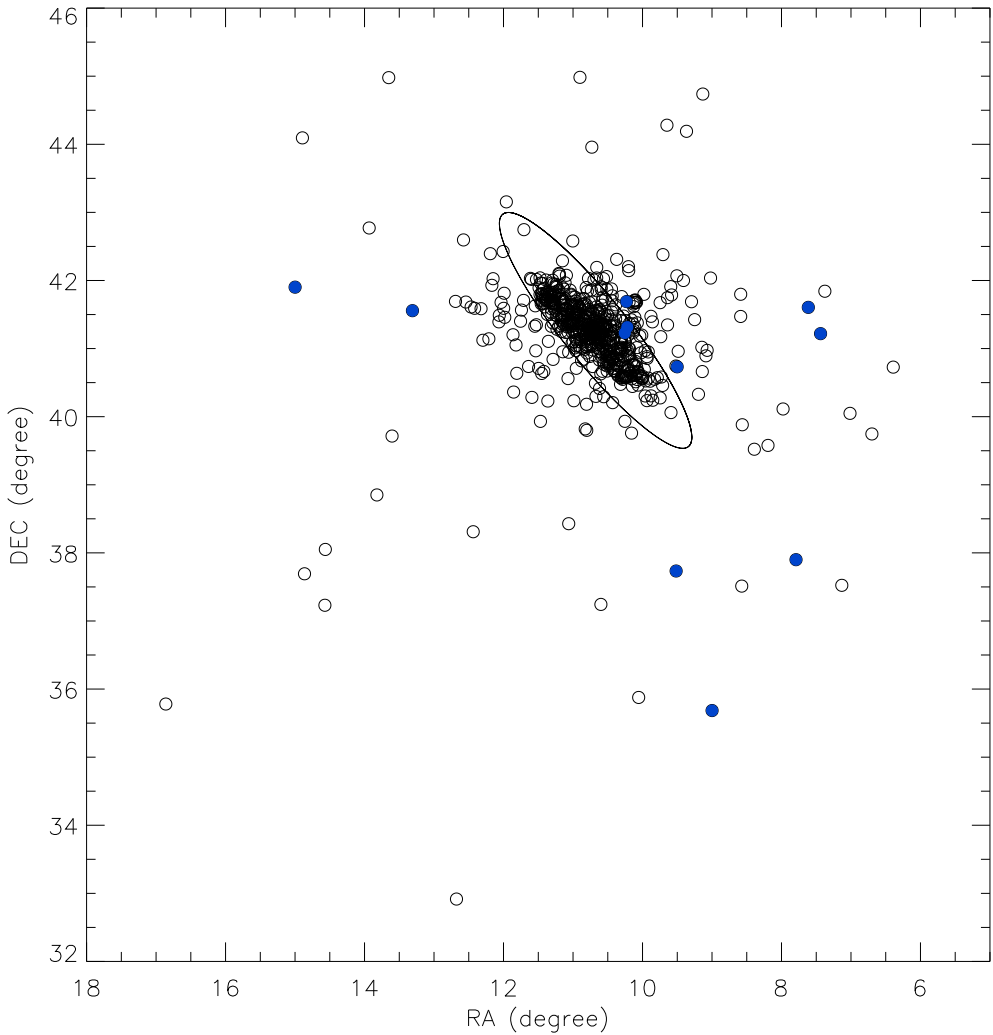}}}
\caption{Spatial distribution of our sample GCs (blue filled
  circles) and all the confirmed GCs from RBC v.4 (open circles). The
  large ellipse is the M31 disk/halo boundary as defined by \citet{rac91}.} 
\label{fig1}
\end{figure}

We show the spatial distribution of our sample GCs and all the
confirmed GCs from RBC v.4 in Figure~\ref{fig1}. The large ellipse is
the M31 disk/halo boundary as defined by \citet{rac91}. 
Note that most of our sample are located in the halo of M31 except
B019 and B023, which are very close to each other with a distance of
$\sim5.5$ arcmin. That's to say, most of GCs in our sample are halo GCs in
M31, which can help us to access the nature of galaxy halo with these clusters.

\section{Observations and data reduction}
\label{obs.sec}

Our Low-resolution spectroscopic observations were all taken at
the 2.16m optical telescope at Xinglong Site, which belongs to
National Astronomical Observatories, Chinese Academy of Sciences (NAOC),
from 10th to 13th September 2010. An OMR (Optomechanics Research Inc.)
spectrograph and a PI 1340${\times}$400 CCD detector were used during
this run with a dispersion of 200 {\AA} mm$^{-1}$, 4.8 {\AA} pixel$^{-1}$,
and a 3.0 \arcsec slit. Exposures of $3\times1800$ seconds were taken with
seeing typically $\sim2.5$ \arcsec. Our spectra cover the wavelength range
of $3500-8100$ {\AA} at 4 {\AA} resolution. All our spectra have 
$S/N \ge 40$.

The spectroscopic data were reduced following standard procedures using the
NOAO Image Reduction and Analysis Facility (IRAF, version 2.11)
software package. The CCD reduction includes bias and flat-field
correction, as well as cosmic-ray removal. Wavelength calibration was
performed based on helium/argon lamps exposed at both the beginning
and the end of the observations each night. Flux calibration of all
spectra was performed based on observations of at least two of the
KPNO spectral standard stars \citep{mass88} per night. The atmospheric
extinction was corrected for using the mean extinction coefficients
measured for Xinglong by the Beijing-Arizona-Taiwan-Connecticut (BATC)
multicolor sky survey (H. J. Yan 1995, priv. comm.).

Before we measure the Lick absorption line indices, the heliocentric radial
velocities $V_r$ were obtained by fitting the abosorption lines of our
spectra with the templates in various radial velocities. The
typical internal velocity errors on a single measure is $\sim 20$ km
s$^{-1}$. The estimated radial velocities $V_r$ with the associated
uncertainties (Col. 2) are listed in Table~\ref{t2.tab}. The published
radial velocities $V_r$ (Col. 3) are also listed for comparisons. It
can be seen that our measurements agree well with those 
listed in RBC v.4. At least, we can not see significant differences
between our measurements and the published values.

\begin{table}[ht!!!]
\small
\centering
\begin{minipage}[]{100mm}
  \caption[]{The radial velocities $V_r$ of our sample GCs as well as
    the previous results.}\label{t2.tab}\end{minipage} 
%%Please Capitalize the First Letter of Each Notional Word in table's caption
\tabcolsep 3mm
\begin{tabular}{lcc}
  \hline\noalign{\smallskip}
  ID & This work & RBC v.4 \\
  \hline\noalign{\smallskip}
MCGC2   & $-586.87\pm39.98$ &             \\
MCGC3   & $-416.46\pm14.01$ &             \\
B514 & $-429.42\pm20.24$ & $-458\pm23$ \\
MCGC5  & $-417.55\pm25.03$ &             \\
B298 & $-648.50\pm16.67$ & $-539\pm12$ \\
H12        & $-412.51\pm33.05$ &             \\
B019       & $-149.83\pm22.91$ & $-224\pm2$  \\
B020       & $-231.87\pm26.48$ & $-351\pm1$  \\
B023       & $-348.44\pm21.30$ & $-451\pm5$  \\
EXT8       & $-104.55\pm7.32$  & $-154\pm30$ \\
B517       & $-267.47\pm20.73$ & $-272\pm54$ \\
  \noalign{\smallskip}\hline
\end{tabular}
\end{table}

\begin{figure}
  \resizebox{\hsize}{!}{\rotatebox{0}{\includegraphics{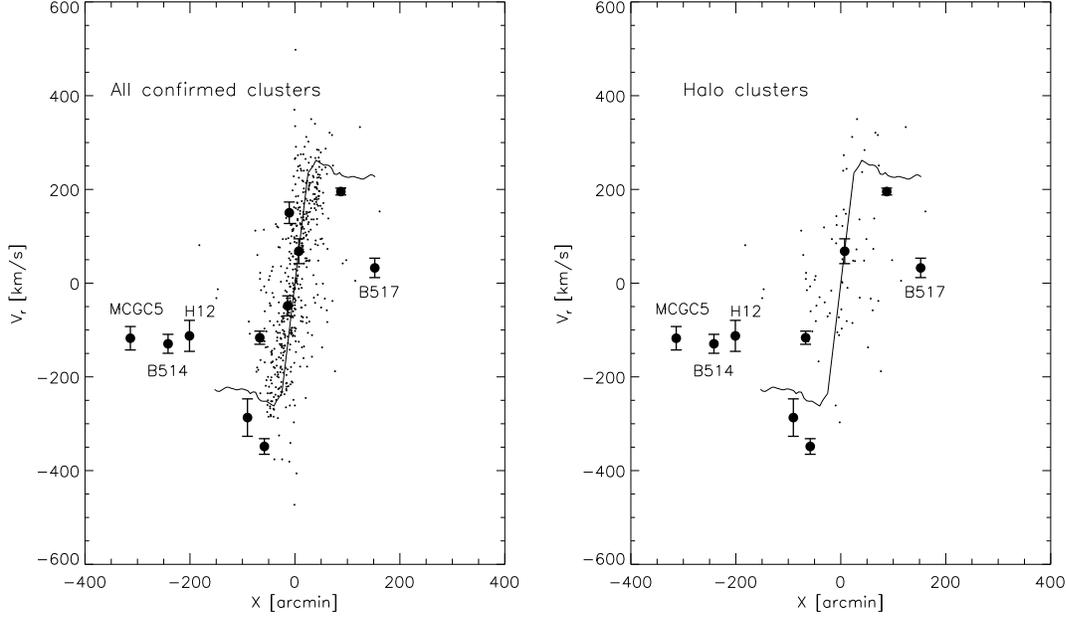}}}
  \caption{The radial velocity $V_r$ (corrected for the systemic
    velocity of M31) as a function of the projected distance along the
    major axis ($X$) in arcmin for all the confirmed clusters
      (left) and the halo clusters (right). The solid line is the HI rotation curve of 
    the galaxy from \citet{cari06}. The filled circles with errors are the
    GCs from our sample while the small points are the velocity from
    RBC v.4 catalogue. It is easy to find out that both the halo
    clusters as well as most of our sample clusters are distinct from HI
    rotation curve of the galaxy, implying that they do not have kinematic
    association with the star forming young disk of the galaxy.} 
  \label{fig2}
\end{figure}

Similar to \citet{gall05,gall06} and \citet{cw09}, we plotted the radial velocity $V_r$ 
(corrected for the systemic velocity of M31) versus the projected distance 
along the major axis ($X$) in Figure~\ref{fig2}. The left panel is for
  all the confirmed clusters while the right panel is for the halo
  clusters which are defined in Figure~\ref{fig1}.
The small points are the published measurements
from RBC v.4 while the filled circles with errors are the measurements in our work.
Since \citet{cari06} calculated the HI rotation curve of M31 out to
$\sim 35$ kpc with the observations results of Effelsberg and Green Bank
100 m telescopes, the HI rotation curve of M31 galaxy were over
plotted in figure~\ref{fig2} with the continuous line. It
can be seen that both the halo clusters and most of our sample
clusters do not follow the disk mean velocity curve very well,
especially for B514, MCGC5, H12 and B517, suggesting that they do not
have kinematic association with the star forming young disk of M31.

Subsequently, all the spectra were shifted to the zero radial
velocity and degraded to the wavelength dependent Lick resolution
with a variable-width Gaussian kernel following the definition of
\cite{wo97}, i.e. 11.5 {\AA} at 4000 {\AA}, 9.2 {\AA} at 4400 {\AA},
8.4 {\AA} at 4900 {\AA}, 8.4 {\AA} at 5400 {\AA}, 9.8 {\AA} at 6000
{\AA}. Thus, we measured all the 25 types of Lick indices strictly
by using the parameters and formulas from \citet{w94} and
\citet{wo97}. The uncertainty of each index was estimated based on
the analytic formulae (11)$-$(18) of \citet{car98}. All the Lick
absorption line indices measurements and $1\sigma$ errors are listed
in Table~\ref{t3.tab}.

\begin{table}[ht!!!]
\small
\centering
\begin{minipage}[]{100mm}
  \caption[]{The Lick absorption line indices of our sample GCs}\label{t3.tab}\end{minipage}
%%Please Capitalize the First Letter of Each Notional Word in table's caption
\tabcolsep 1mm
\begin{tabular}{lccccccccccc}
  \hline\noalign{\smallskip}
  Indices & MCGC2 & MCGC3 & B514 & MCGC5 & B298 & H12 & B019 & B020 &  B023 & EXT08 &  B517\\
  \hline\noalign{\smallskip}
$\rm H\delta_A$ (\AA)&$-2.297$&$ 3.038$&$ 2.526$&$ 3.450$&$ 4.520$&$ 3.331$&$ 1.640$&$ 1.592$&$ 0.719$&$ 4.054$&$ 2.797$\\
    error&$ 0.351$&$ 0.270$&$ 0.315$&$ 0.294$&$ 0.532$&$ 0.249$&$ 0.309$&$ 0.347$&$ 0.228$&$ 0.322$&$ 0.258$\\
$\rm H\delta_F$ (\AA)&$ 0.154$&$ 2.258$&$ 1.916$&$ 2.839$&$ 2.775$&$ 1.977$&$ 0.661$&$ 1.136$&$ 0.589$&$ 2.497$&$-0.358$\\
    error&$ 0.237$&$ 0.260$&$ 0.256$&$ 0.307$&$ 0.362$&$ 0.213$&$ 0.233$&$ 0.256$&$ 0.174$&$ 0.215$&$ 0.204$\\
$\rm  CN1$ (mag)&$ 0.037$&$-0.084$&$-0.095$&$-0.137$&$-0.104$&$-0.069$&$-0.039$&$-0.056$&$ 0.031$&$-0.142$&$-0.097$\\
    error&$ 0.011$&$ 0.008$&$ 0.007$&$ 0.009$&$ 0.011$&$ 0.007$&$ 0.007$&$ 0.010$&$ 0.005$&$ 0.008$&$ 0.006$\\
$\rm  CN2$ (mag)&$ 0.097$&$ 0.015$&$-0.013$&$-0.060$&$ 0.029$&$-0.023$&$ 0.033$&$ 0.013$&$ 0.082$&$-0.066$&$-0.022$\\
    error&$ 0.015$&$ 0.022$&$ 0.015$&$ 0.019$&$ 0.024$&$ 0.017$&$ 0.014$&$ 0.012$&$ 0.011$&$ 0.020$&$ 0.011$\\
$\rm Ca4227$ (\AA)&$ 0.358$&$ 0.383$&$ 0.000$&$ 0.119$&$ 0.340$&$ 0.582$&$ 0.388$&$ 0.170$&$ 0.230$&$ 0.127$&$ 0.085$\\
    error&$ 0.219$&$ 0.093$&$ 0.040$&$ 0.110$&$ 0.118$&$ 0.148$&$ 0.097$&$ 0.076$&$ 0.074$&$ 0.047$&$ 0.156$\\
$\rm  G4300$ (\AA)&$ 1.416$&$ 1.577$&$ 1.381$&$ 3.088$&$ 0.954$&$ 1.221$&$ 3.117$&$ 2.871$&$ 3.534$&$ 0.272$&$ 1.901$\\
    error&$ 0.482$&$ 0.323$&$ 0.225$&$ 0.396$&$ 0.254$&$ 0.393$&$ 0.319$&$ 0.448$&$ 0.414$&$ 0.225$&$ 0.172$\\
$\rm H\gamma_A$ (\AA)&$-0.747$&$ 1.064$&$ 0.954$&$-1.626$&$ 2.603$&$-0.834$&$-4.478$&$-6.141$&$-5.271$&$ 1.746$&$-0.215$\\
    error&$ 0.489$&$ 0.277$&$ 0.249$&$ 0.284$&$ 0.206$&$ 0.339$&$ 0.394$&$ 0.381$&$ 0.400$&$ 0.288$&$ 0.239$\\
$\rm H\gamma_F$ (\AA)&$ 1.747$&$ 1.364$&$ 1.307$&$ 0.339$&$ 1.346$&$ 1.022$&$-0.321$&$-0.788$&$-0.056$&$ 1.924$&$ 0.923$\\
    error&$ 0.254$&$ 0.199$&$ 0.167$&$ 0.252$&$ 0.133$&$ 0.257$&$ 0.157$&$ 0.179$&$ 0.168$&$ 0.181$&$ 0.179$\\
$\rm Fe4383$ (\AA)&$-0.202$&$-0.202$&$-0.520$&$-0.654$&$ 0.240$&$ 0.816$&$ 1.957$&$ 2.439$&$ 3.115$&$ 0.357$&$ 0.244$\\
    error&$ 0.543$&$ 0.406$&$ 0.286$&$ 0.240$&$ 0.566$&$ 0.241$&$ 0.335$&$ 0.375$&$ 0.379$&$ 0.199$&$ 0.301$\\
$\rm Ca4455$ (\AA)&$ 0.321$&$ 0.399$&$ 0.055$&$ 0.243$&$ 1.309$&$ 0.391$&$ 0.530$&$ 0.291$&$ 0.601$&$ 0.118$&$ 0.810$\\
    error&$ 0.252$&$ 0.224$&$ 0.078$&$ 0.078$&$ 0.260$&$ 0.121$&$ 0.183$&$ 0.099$&$ 0.127$&$ 0.030$&$ 0.219$\\
$\rm Fe4531$ (\AA)&$ 0.169$&$ 0.300$&$ 0.287$&$ 0.823$&$ 0.298$&$-0.642$&$ 1.929$&$ 1.711$&$ 1.582$&$ 0.281$&$ 2.319$\\
    error&$ 0.545$&$ 0.212$&$ 0.120$&$ 0.223$&$ 0.208$&$ 0.167$&$ 0.232$&$ 0.348$&$ 0.155$&$ 0.111$&$ 0.237$\\
$\rm Fe4668$ (\AA)&$-1.582$&$-0.848$&$ 2.038$&$-1.400$&$ 0.662$&$-2.585$&$ 2.566$&$ 1.162$&$ 0.848$&$-0.498$&$ 1.024$\\
    error&$ 0.405$&$ 0.289$&$ 0.251$&$ 0.336$&$ 0.433$&$ 0.335$&$ 0.347$&$ 0.181$&$ 0.214$&$ 0.080$&$ 0.311$\\
$\rm  H\beta$ (\AA)&$ 2.212$&$ 1.794$&$ 2.250$&$ 2.269$&$ 2.308$&$ 3.082$&$ 2.037$&$ 1.791$&$ 1.526$&$ 2.583$&$ 3.148$\\
    error&$ 0.187$&$ 0.209$&$ 0.211$&$ 0.243$&$ 0.254$&$ 0.305$&$ 0.336$&$ 0.242$&$ 0.214$&$ 0.219$&$ 0.290$\\
$\rm Fe5015$ (\AA)&$-0.557$&$ 1.044$&$ 1.506$&$ 1.305$&$-0.388$&$ 2.170$&$ 3.209$&$ 3.281$&$ 2.242$&$ 0.342$&$ 1.903$\\
    error&$ 0.386$&$ 0.216$&$ 0.144$&$ 0.198$&$ 0.137$&$ 0.359$&$ 0.384$&$ 0.304$&$ 0.167$&$ 0.076$&$ 0.271$\\
$\rm  Mg1$ (mag)&$ 0.054$&$ 0.004$&$-0.007$&$-0.003$&$ 0.011$&$ 0.014$&$ 0.034$&$ 0.017$&$ 0.032$&$ 0.015$&$ 0.001$\\
    error&$ 0.005$&$ 0.001$&$ 0.001$&$ 0.002$&$ 0.002$&$ 0.002$&$ 0.002$&$ 0.003$&$ 0.002$&$ 0.001$&$ 0.002$\\
$\rm  Mg2$ (mag)&$ 0.053$&$ 0.027$&$ 0.028$&$ 0.027$&$ 0.033$&$ 0.036$&$ 0.121$&$ 0.102$&$ 0.119$&$ 0.004$&$ 0.044$\\
    error&$ 0.004$&$ 0.004$&$ 0.003$&$ 0.004$&$ 0.002$&$ 0.002$&$ 0.008$&$ 0.006$&$ 0.006$&$ 0.002$&$ 0.004$\\
$\rm    Mgb$ (\AA)&$ 0.506$&$ 0.415$&$ 0.518$&$ 1.370$&$ 0.109$&$ 0.731$&$ 2.392$&$ 2.178$&$ 1.927$&$ 0.088$&$ 0.962$\\
    error&$ 0.165$&$ 0.140$&$ 0.098$&$ 0.156$&$ 0.093$&$ 0.105$&$ 0.241$&$ 0.214$&$ 0.163$&$ 0.063$&$ 0.164$\\
$\rm Fe5270$ (\AA)&$ 0.202$&$ 0.409$&$ 0.919$&$ 0.339$&$-0.185$&$ 0.148$&$ 1.345$&$ 1.900$&$ 1.526$&$ 0.225$&$ 1.618$\\
    error&$ 0.270$&$ 0.086$&$ 0.140$&$ 0.111$&$ 0.137$&$ 0.103$&$ 0.184$&$ 0.252$&$ 0.205$&$ 0.043$&$ 0.194$\\
$\rm Fe5335$ (\AA)&$-0.391$&$ 0.529$&$ 0.166$&$ 1.082$&$ 0.703$&$-0.221$&$ 1.014$&$ 0.979$&$ 1.187$&$ 0.439$&$ 0.730$\\
    error&$ 0.217$&$ 0.093$&$ 0.091$&$ 0.187$&$ 0.125$&$ 0.101$&$ 0.190$&$ 0.195$&$ 0.177$&$ 0.074$&$ 0.177$\\
$\rm Fe5406$ (\AA)&$ 0.481$&$-0.388$&$-0.061$&$ 0.282$&$-0.235$&$ 0.234$&$ 0.993$&$ 0.557$&$ 0.730$&$ 0.224$&$ 0.108$\\
    error&$ 0.268$&$ 0.193$&$ 0.091$&$ 0.108$&$ 0.144$&$ 0.115$&$ 0.166$&$ 0.092$&$ 0.113$&$ 0.059$&$ 0.101$\\
$\rm Fe5709$ (\AA)&$-0.045$&$ 0.021$&$ 0.008$&$ 0.325$&$-0.416$&$ 0.286$&$ 0.351$&$ 0.021$&$ 0.483$&$ 0.050$&$ 0.025$\\
    error&$ 0.082$&$ 0.041$&$ 0.020$&$ 0.061$&$ 0.103$&$ 0.107$&$ 0.075$&$ 0.078$&$ 0.087$&$ 0.028$&$ 0.111$\\
$\rm Fe5782$ (\AA)&$ 0.334$&$ 0.142$&$ 0.177$&$ 0.239$&$ 0.275$&$-0.143$&$ 0.311$&$ 0.186$&$ 0.512$&$ 0.103$&$-0.128$\\
    error&$ 0.105$&$ 0.028$&$ 0.053$&$ 0.062$&$ 0.094$&$ 0.077$&$ 0.064$&$ 0.053$&$ 0.097$&$ 0.029$&$ 0.086$\\
$\rm    NaD$ (\AA)&$ 1.025$&$ 1.492$&$ 1.175$&$ 1.446$&$ 1.663$&$ 1.559$&$ 3.491$&$ 2.490$&$ 3.642$&$ 0.744$&$ 0.215$\\
    error&$ 0.113$&$ 0.121$&$ 0.094$&$ 0.164$&$ 0.192$&$ 0.150$&$ 0.399$&$ 0.254$&$ 0.380$&$ 0.083$&$ 0.070$\\
$\rm TiO1$ (mag)&$ 0.038$&$ 0.004$&$ 0.007$&$ 0.015$&$-0.008$&$ 0.009$&$ 0.036$&$ 0.024$&$ 0.046$&$ 0.012$&$ 0.002$\\
    error&$ 0.003$&$ 0.002$&$ 0.002$&$ 0.001$&$ 0.003$&$ 0.002$&$ 0.001$&$ 0.002$&$ 0.001$&$ 0.001$&$ 0.004$\\
$\rm TiO2$ (mag)&$-0.009$&$ 0.011$&$ 0.002$&$ 0.016$&$ 0.019$&$-0.024$&$ 0.062$&$ 0.050$&$ 0.064$&$ 0.009$&$ 0.015$\\
    error&$ 0.003$&$ 0.001$&$ 0.001$&$ 0.001$&$ 0.002$&$ 0.002$&$ 0.001$&$ 0.001$&$ 0.001$&$ 0.001$&$ 0.002$\\
  \noalign{\smallskip}\hline
\end{tabular}
\end{table}

\begin{figure*}
\resizebox{\hsize}{!}{\rotatebox{0}{\includegraphics{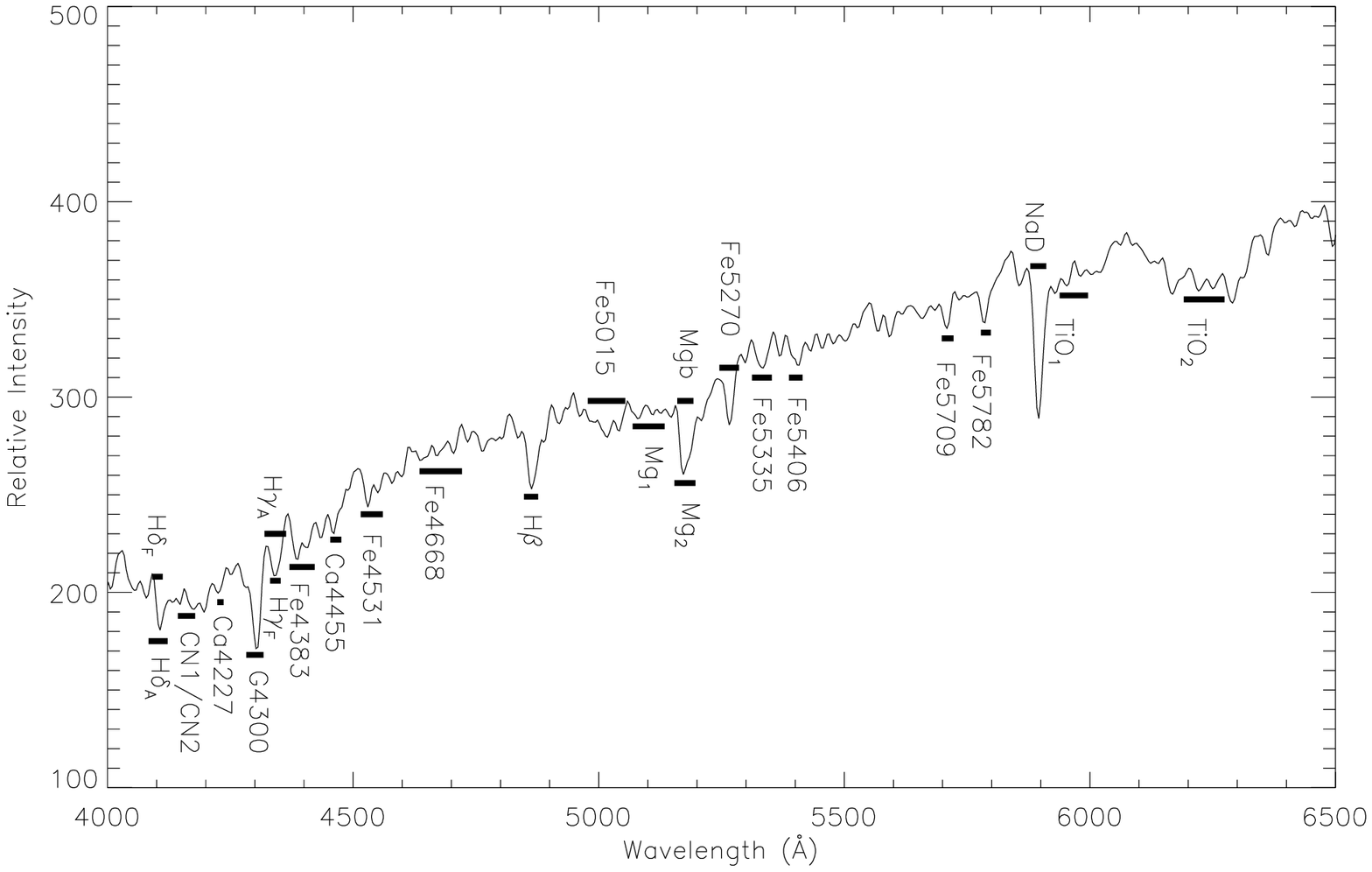}}}
\caption{Spectrum of GC B023 in our sample, with the index bandpasses of
  all the absorption Lick indices defined in \citet{w94} and \citet{wo97}
  marked. As we can see, the index bandpasses for CN1 and CN2 are the
  same and the only difference is the pseudocontinua coverage.}
  \label{fig3}
\end{figure*}

As an example, Figure \ref{fig3} shows the reduced spectroscopy of
our sample GC B023, 
with all the Lick absorption line indices bandpasses marked. The
spectrum has been degraded and shifted to the zero radial velocity
as described above. Actually, from the definitions of line indices
of CN1 and CN2 \citep{w94,wo97}, we find that the index bandpasses of
them are totally the same and the only difference is the
pseudocontinua coverage.

A simple way to estimate the metallicity is by calculating it from the
combination of absorption line indices Mg and Fe. \citet{gall09} provide the method to
measure the metallicity from [MgFe], which is defined as [MgFe] $= \rm
\sqrt{Mg{\it b}\langle Fe \rangle}$, with $\rm \langle Fe \rangle =$
(Fe5270+Fe5335)/2. Thus, the metallicity can be derived from the formula
below,
\begin{equation}
  \rm [Fe/H]_{[MgFe]}=-2.563+1.119[MgFe]-0.106[MgFe]^2\pm0.15.
  \label{eq1}
\end{equation}

The uncertainty of the $\rm [Fe/H]_{[MgFe]}$ was estimated with the
equation in the following,
\begin{equation}
  \rm
  % \sigma_{[Fe/H]}^2=1.119^2\sigma_{[MgFe]}^2+0.106^2\sigma_{[MgFe]}^4
  \sigma_{[Fe/H]}^2=1.119^2\sigma_{[MgFe]}^2+4\times0.106^2[MgFe]^2\sigma_{[MgFe]}^2.
  \label{eq2}
\end{equation}
All metallicity $\rm [Fe/H]_{[MgFe]}$ derived from [MgFe] and the
associated uncertainty determinations are listed in Col. (2) of
Table~\ref{t4.tab}. It is obvious that all the metallicity derived
from the line index [MgFe] agree well with those from the model
fitting method. 

\section{Fitting, analysis and results}
\label{fit.sec}

\subsection{Model description}

\citet{tmb} provided stellar population models including Lick
absorption line indices for various elemental-abundance ratios, covering
ages from 1 to 15 Gyr and metallicities from 1/200 to $3.5 \times$
solar abundance. These models are based on the standard models of
\citet{mar98}, with input stellar evolutionary tracks from
\citet{ccc97} and \citet{bo97} and a \citet{sal} stellar initial mass
function. \citet{tmk} improved the models by including higher-order
Balmer absorption-line indices. They found that these Balmer indices
are very sensitive to changes in the $\alpha$/Fe ratio for supersolar
metallicities. The latest stellar population model for Lick
absorption-line indices \citep{tmj} is an improvement on \citet{tmb}
and \citet{tmk}. They were derived from the MILES stellar library,
which provides a higher spectral resolution appropriate for MILES and
Sloan Digital Sky Survey (SDSS) spectroscopy, as well as flux
calibration. The models cover ages from 0.1 to 15 Gyr, [Fe/H]
from $-2.25$ to 0.67 dex, and [$\alpha$/Fe] from $-0.3$ to 0.5 dex.
In our work, we fitted our absorption indices based on the models of
\citet{tmj}, by using the two sets of stellar evolutionary
tracks provided, i.e., \citet{ccc97} and Padova.

\subsection{Fitting with stellar population models and the results}

As \citet{cw09} demonstrated that the $\chi^2-$minimization
method for many diagnostic lines are more reliable to extract the
ages than the two absorption line indices diagram plot
method. Furthermore, we have measured 25 
different types of Lick line indices listed in Table~\ref{t3.tab},
all of which were used for the fitting procedure, then the
results should be much more reliable and accurate.
Since \citet{tmj} provide only 20 ages, 6 [Fe/H] values, and 4
[$\alpha$/Fe], it is necessary to interpolate the original models
to the higher-resolution models for our needs. We carried out the
cubic spline interpolations, using equal step lengths, to obtain a grid
of 150 ages from 0.1 to 15 Gyr, 31 [Fe/H] values from $-2.25$ to 0.67
dex, and 51 [$\alpha$/Fe] from $-0.3$ to 0.5 dex, which makes the
model more accurate and more helpful for our following statistics. 
Therefore, the ages ($t$), metallicities [Fe/H], and
[$\alpha$/Fe] were determined at the same time by comparing the
interpolated stellar population models with the spectral-energy
distributions from our photometry by employing the $\chi^2-$minimization    
method, i.e.,
\begin{equation}
  \chi^2_{\rm min}(t,{\rm Fe/H},\alpha/{\rm Fe})={\rm
    min}\left[\sum_{i=1}^{25}\left({\frac{L_{\lambda_i}^{\rm
            obs}-L_{\lambda_i}^{\rm mod}}{\sigma_i}}\right)^2\right],
\label{eq3}
\end{equation}
where $L_{\lambda_i}^{\rm mod}(t,{\rm Fe/H},\alpha$/Fe) is the $i^{\rm th}$
Lick line index in the stellar population model for age $t$, metallicity
[Fe/H], and [$\alpha$/Fe], while $L_{\lambda_i}^{\rm obs}$ represents the
observed Lick line indices from our measurements and the errors
estimated in our fitting are given as follows,
\begin{equation}
\sigma_i^{2}=\sigma_{{\rm obs},i}^{2}+\sigma_{{\rm mod},i}^{2}.
\label{eq4}
\end{equation}
Here, $\sigma_{{\rm obs},i}$ is the observational uncertainty and
$\sigma_{{\rm mod},i}$ is the uncertainty associated with the models of
\citet{tmj}. We combined the two uncertainties together in our fitting.

\begin{table}
\bc
\begin{minipage}[]{100mm}
  \caption[]{The $\chi^2-$minimization fitting results using \citet{tmj} models
    with \citet{ccc97} and Padova stellar tracks, respectively.}\label{t4.tab}
\end{minipage}
\small
\tabcolsep 1mm
 \begin{tabular}{lcccrccr}
  \hline\noalign{\smallskip}
   ID & $\rm [Fe/H]_{[MgFe]}$ &$\rm [Fe/H]_{Cas}$ &$\rm {Age_{Cas}
  (Gyr)} $& $\rm [\alpha/Fe]_{Cas} $& $\rm [Fe/H]_{Pad}$ &$\rm
{Age_{Pad} (Gyr)} $& $\rm [\alpha/Fe]_{Pad} $ \\
  \hline\noalign{\smallskip}
  MCGC2  &   $ -2.32\pm0.28 $  &$ -1.53^{+0.28}_{-0.18}$ &
  $13.60^{+0.40}_{-0.50}$ &$0.46^{+0.02}_{-0.76}$&
  $-1.44^{+0.19}_{-0.18}$ & $13.60^{+0.50}_{-1.00}$ &
  $0.50^{+0.00}_{-0.80}$\\
  MCGC3    &   $ -2.09\pm0.13 $ & $-1.80^{+0.09}_{-0.18}$  &
  $13.60^{+0.50}_{-2.30}$ &$0.50^{+0.00}_{-0.47}$ &
  $-1.80^{+0.18}_{-0.18}$ & $13.60^{+0.80}_{-2.90}$ &
  $0.50^{+0.00}_{-0.50}$\\
  B514    &  $ -2.00\pm0.12 $  &$ -1.89^{+0.09}_{-0.18}$ &
  $13.60^{+0.40}_{-0.60}$ & $0.50^{+0.00}_{-0.38}$ &
  $-1.89^{+0.18}_{-0.18}$ & $13.60^{+1.30}_{-2.70}$ &
  $0.50^{+0.00}_{-0.35}$\\
  MCGC5    &  $  -1.56\pm0.18 $ & $ -1.53^{+0.18}_{-0.18} $ &
  $13.60^{+1.40}_{-0.60}$ &$0.50^{+0.00}_{-0.35}$ &
  $-1.44^{+0.09}_{-0.18}$ & $13.60^{+0.60}_{-2.40}$ &
  $0.50^{+0.00}_{-0.29}$\\
  B298    &   $ -2.38\pm0.13 $ &$ -2.07^{+0.18}_{-0.09}$  &
  $13.60^{+0.50}_{-3.00}$ &$0.50^{+0.00}_{-0.65}$ &
  $-2.07^{+0.18}_{-0.18}$ & $13.60^{+0.60}_{-3.40}$ &
  $0.50^{+0.00}_{-0.62}$\\
  H12    &   $ -2.38\pm0.24 $ & $-1.80^{+0.18}_{-0.18}$  &
  $13.60^{+1.40}_{-1.70}$ & $0.50^{+0.00}_{-0.26}$ &
  $-1.71^{+0.18}_{-0.18}$ & $14.80^{+0.20}_{-4.20}$ &
  $0.50^{+0.00}_{-0.20}$\\
  B019   &   $ -0.98\pm0.26$ & $ -0.74^{+0.10} _{-0.20}$ &
  $13.50^{+0.80}_{-2.70}$ &$0.48^{+0.02}_{-0.33}$
  &$-0.53^{+0.10}_{-0.10}$ & $7.70^{+6.30}_{-0.80}$ &
  $0.48^{+0.02}_{-0.36}$ \\
  B020   &   $ -0.91\pm0.26$ & $-0.94^{+0.20}_{-0.10} $ &
  $13.70^{+0.30}_{-0.50}$ &$0.44^{+0.06}_{-0.41}$
  &$-0.94^{+0.10}_{-0.10}$ & $13.60^{+0.30}_{-0.30}$ &
  $0.50^{+0.00}_{-0.35}$\\
  B023   &   $ -1.03\pm0.21$  &$-0.84^{+0.10}_{-0.10}$ &
  $13.60^{+0.50}_{-0.60}$&$0.34^{+0.16}_{-0.37}$
  &$-0.74^{+0.10}_{-0.20}$ &$13.20^{+0.20}_{-1.50}$ &
  $0.40^{+0.10}_{-0.34}$\\
  EXT8 &   $ -2.38\pm0.09 $ & $-2.07^{+0.09}_{-0.18}$  &
  $13.50^{+0.60}_{-4.30}$ &$0.50^{+0.00}_{-0.80}$ &
  $-2.07^{+0.09}_{-0.18}$ & $13.50^{+1.40}_{-4.50}$ &
  $0.50^{+0.00}_{-0.80}$\\
  B517   &  $  -1.49\pm0.20 $ & $-1.53^{+0.09}_{-0.18}$  &
  $13.60^{+1.40}_{-0.60}$ &$0.00^{+0.50}_{-0.30}$ &
  $-1.53^{+0.18}_{-0.09}$ & $13.60^{+0.60}_{-1.70}$ &
  $0.00^{+0.50}_{-0.30}$\\
  \noalign{\smallskip}\hline
\end{tabular}
\ec
\tablecomments{0.86\textwidth}{Cas: fitting with the model evolutionary tracks
  of \citet{ccc97}; Pad:  fitting with the model evolutionary tracks
  of Padova, respectively.}
\end{table}

From Table \ref{t4.tab}, we found that either the [Fe/H] or
the ages or the [$\alpha$/Fe] derived from the \citet{ccc97} or from
Padova tracks of the models are basically the same, suggesting that
our fitting results are consistent with each other. Moreover, the ages
constrained in our work are in 
good agreement with those previous work in Table~\ref{t4.tab},
implying that our fitting method is reliable. Besides, it is worth 
noting that all of our sample GCs, most of which are located in the galaxy
halo, are older than 10 Gyr, indicating that these clusters formed at the very 
beginning of the galaxy formation. We also find that the
metallicity derived from the absorption-line index [MgFe] (in Col. 2) consists
with metallicity fitted with two different tracks of the model (in
Cols. 3 and 6). Previously, \citet{gall05} estimated the
metallicity of B514 with [Fe/H]$=-1.8\pm0.3$,  
and RBC v.4 lists the metallicity for B514 [Fe/H]$=-2.06\pm0.16$, 
for B298 [Fe/H]$=-1.78\pm0.22$, for B019 [Fe/H]$=-0.74\pm0.15$, 
for B020 [Fe/H]$=-0.83\pm0.07$, for B023 [Fe/H]$=-0.91\pm0.14$.  
All these previous measurements are in good agreement with our estimates 
in Table~\ref{t4.tab}, indicating our method and results are reliable.
For consistency, in the subsequent analysis, we adopted the ages and
metallicity from \citet{tmj} + \citet{ccc97} track of the predictive model.

\subsection{Metallicity Properties of Outer Halo}
\label{md.sec}

The metal abundance is one of the most important properties of star
clusters to understand the formation and enrichment processes of their
host galaxy. For instance, the halo stars and clusters should
feature large-scale metallicity gradients if the enrichment timescale
is shorter than the collapse time, which may be due to the galaxy
formation as a consequence of a monolithic, dissipative, and rapid
collapse of a single massive, nearly spherical, spinning gas cloud
\citep{eggen62,bh00}. On the other hand, \citet{sz78} presented a
chaotic scheme for early galactic evolution, when the loosely bound
pre-enriched fragments merge with the protogalaxy during a very long
period of time, in which case a more homogeneous metallicity
distribution should develop. Most galaxies are thought to have formed
through a combination of these scenarios. 

A lot of previous work attempted to find the clues of formation and
evolution for M31 galaxy through studying the metallicity distribution
of its globular cluster system. \citet{hbk91,ash93,bh00,per02,fan08}  
found that the metal-rich GCs are statistically more concentrated toward 
the center of the galaxy, while their metal-poor counterparts are more 
spatially extended in the halo. Furthermore, there are also many work
to find out whether or not a radial metallicity gradient exists for
M31 star cluster system. \citet{van69,hsv82} showed that 
there is little or no evidence for a general radial metallicity gradient 
for GCs within a radius of 50 arcmin. However, studies including 
\citet{hbk91,per02,fan08} support the possible
existence of a radial metallicity gradient for the metal-poor M31 GCs,
although the slope is not very significant. \citet{per02} suggest that
the gradients is $-0.017$ and $-0.015$ dex arcmin$^{-1}$ for the full
sample and inner metal-poor clusters. More recently, \citet{fan08}
found that the slope is $-0.006$ and $-0.007$ dex arcmin$^{-1}$ for
the metal-poor subsample and whole sample while the slope approches zero 
for the metal-rich subsample. Nevertheless, all these studies are based 
on GCs that are located relatively close to the center of the galaxy, 
usually at projected radii of less than 100 arcmin. In our work, we extended 
the radial coverage to a radius of $r_{\rm p} \sim117$ kpc, which corresponds 
to $\sim510$ arcmin, to check if the previous findings are correct at a 
much larger distance from the galaxy center.

For the purpose of better investigating the metallicity
distribution/spatial gradient, we enlarged the metallicity sample
by merging the metallicity of our measurements 
with the published spectroscopic metallicity from 
\citet{hbk91,bh00,per02,gall09,cw11} together with those from CMD
fittings \citet{mac06,mac07,mac10} . For the published data, if the
metallicity from different work overlapped with the other, the small
smaller associated uncertainty data will superseded the larger
one and the spectroscopic data will superseded the one derived from
CMD fitting. In total, we have a metallicity sample of 384 entries.

\begin{figure}
\resizebox{\hsize}{!}{\rotatebox{0}{\includegraphics{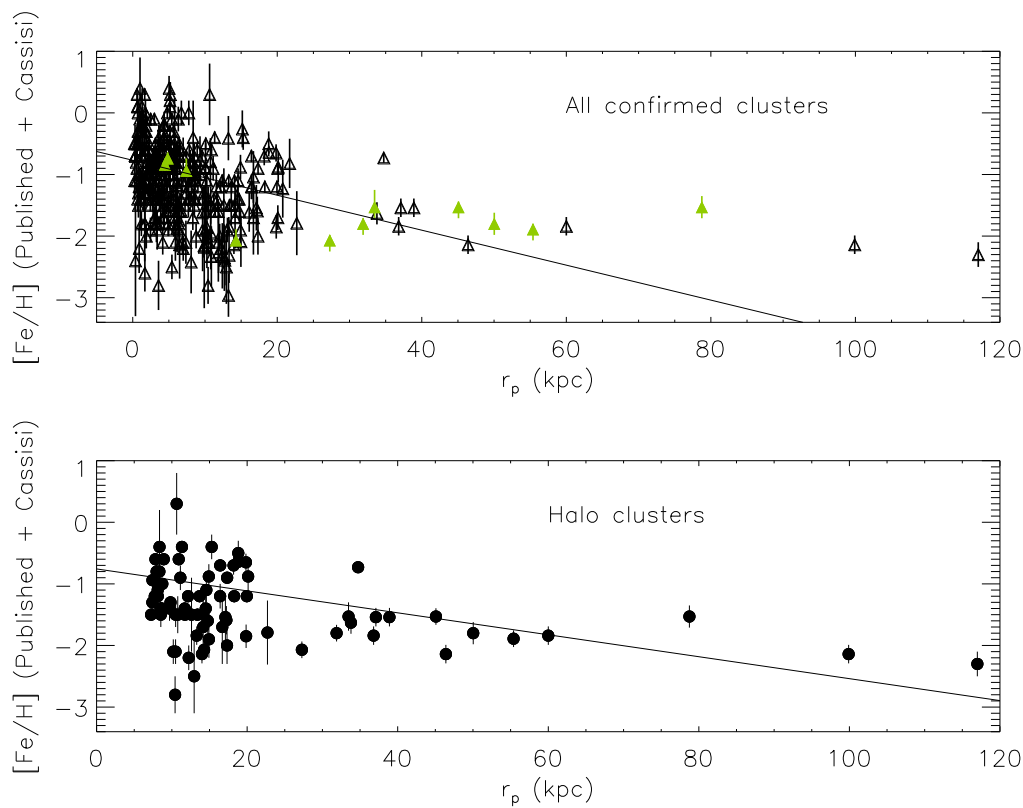}}}
\caption{[Fe/H] versus projected radius from the galaxy center
  for M31 GCs. The solid line refers to a linear fit to all
  the data. {\it Top}: All the confirmed clusters. The open
  triangles with errorbars represent published metallicities while the
  green filled triangles with errorbars are our measurements. {\it Bottom}:
  Halo clusters only. All the clusters are marked with filled circles.} 
\label{fig4}
\end{figure}

Figure~\ref{fig4} shows the metallicity as a function of projected radius 
from the galaxy center for all M31 confirmed clusters (Top) and the
halo clusters only (Bottom) in the unit of kpc. In the top panel, open
triangles with errorbars represent the 
spectroscopic metallicities from the published measurements 
of \citet{hbk91,bh00,per02,gall09,cw11} as well as the metallicities
from CMD fittings \citet{mac06,mac07,mac10} while the green filled 
triangles with errorbars are our spectroscopic measurements. 
The solid line is a linear fit to all the data points, with a slope of
$-0.028\pm0.001$ dex kpc$^{-1}$, responding to $-0.007$ dex
arcmin$^{-1}$.
The fit results are similar to those given by the previous works
\citep[see, ][]{per02,fan08}, which are based on the cluster sample
within a projected radii $r_{\rm p}<100$ arcmin ($\sim23$ kpc). Thus,
our work tentatively supports the notion that a radial metallicity
gradient may exist out to a projected radius of $\sim117$ kpc by
merging the published metallicities. In other words, we updated the
results with the new sample extended to M31's most remote outer
halo. Since the aim of our work is to study the nature of M31 halo, we
would like to foucs on the metallicity gradient of halo clusters. 
In the bottom panel, we only plot the halo clusters, which are defined in
Figure~\ref{fig1}. A least-squares 
fitting yields the slope of $-0.018\pm0.001$ dex kpc$^{-1}$. Therefore,
it can be seen that the metallicity gradient seems to exist for the
halo clusters, although it is not significant.

\begin{figure}
\resizebox{\hsize}{!}{\rotatebox{0}{\includegraphics{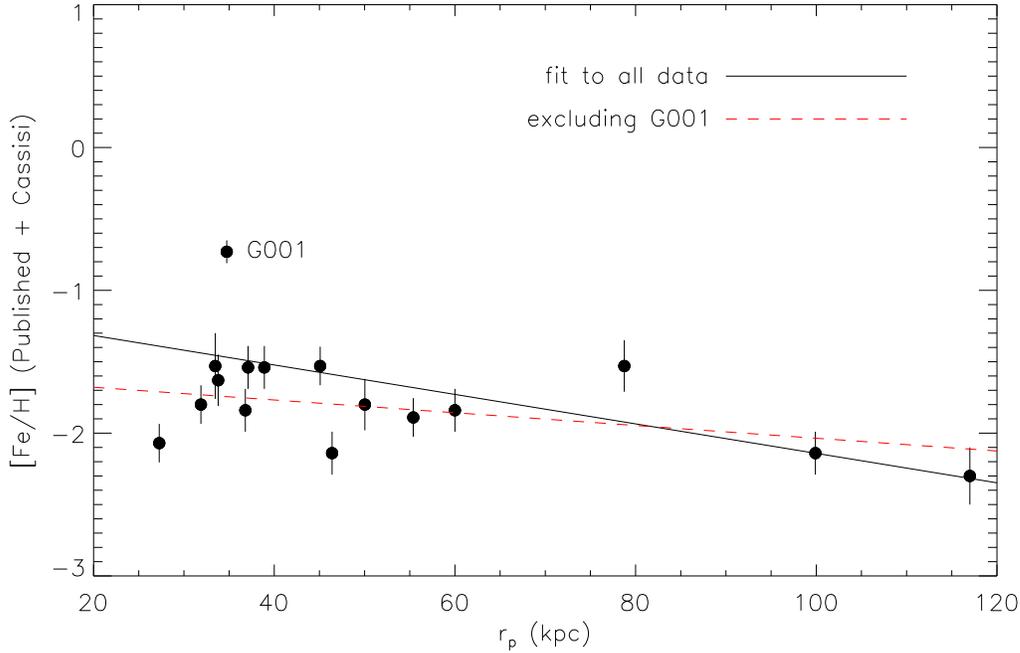}}}
\caption{[Fe/H] from versus projected radius from the galaxy center
  for the outer halo GCs, whose $r_{\rm p} >$ 25 kpc from the center
  of the galaxy. The slope of the fitting for all the data is
  $-0.010\pm0.002$ dex kpc$^{-1}$. However, if G001 is excluded, the
  slope turns out to be $-0.004\pm0.002$ dex kpc$^{-1}$.}  
\label{fig5}
\end{figure}

Furthermore, It is noted that in Figure~\ref{fig4}, the metallicity
gradient for the clusters located in the outer halo with $r_{\rm p} >$
25 kpc is not significant. Thus, we show the metallicity as a function
of projected radius for only the outer halo clusters with $r_{\rm p} >$
25 kpc in Figure~\ref{fig5}. A least-squares linear fitting for all the data
shows the slope is $-0.010\pm0.002$ dex kpc$^{-1}$ (the solid line).
However, if G001 is excluded, the slope turns out to be
$-0.004\pm0.002$ dex kpc$^{-1}$ (the red dashed line), which is
much shallower than that in Figure~\ref{fig4}. It may suggest that the
metallicity gradient is not significant for the outer halo clusters in
M31. Very recently, \citet{h11} investigated the metallicity
gradient for 15 halo CGs to $r_{\rm p}=$117 kpc with the metallicity
derived from the CMD fittings \citet{mac06,mac07,mac10} and the
authors found that the metallicity gradient becomes not significant if
one halo GC H14 is excluded in their Figure~6. We found that our
result is consistent with the previous finding of \citet{h11}. 

\section{Discussions and Summary}
\label{sum.sec}

In our work, we carried out the spectroscopic observations of 11 
confirmed globular clusters of M31 with the OMR spectrograph and a PI
1340${\times}$400 CCD detector on 2.16 m telescope at Xinglong site of
NAOC from 10th to 13th September of 2010. Since our aim is to study
the nature of the halo of M31, we selected the bright confirmed
clusters, 9 of which are located in the halo, out to a projected
radius of 78.75 kpc from the galactic center.  

For all our sample clusters, we measured all types of the Lick
absorption-line indices \citep[see the definitions in,][]{w94,wo97} as
well as the radial velocities.  
We found that most GCs of our sample are distinct from the HI rotation
curve of M31 galaxy, especially for B514, MCGC5, H12 and B517,
suggesting that most of our sample clusters do not have kinematic
association with the star forming young disk of the galaxy.

Since \citet{cw09} demonstrated that the $\chi^2-$minimization method
for many diagnostic lines are more reliable for extracting the ages
than the line indices diagram plot, in our wok we applied the 
$\chi^2-$minimization method to fit the line indices with the updated
stellar population model \citet{tmj} with two different tracks Cassisi
and Padova, separately. The fitting results show that all 
our sample clusters are older than 10 Gyr and most of them are
metal-poor ($-0.91 \le$ [Fe/H] $\le -2.38$ dex). 

In order to enlarge our sample, we merged the spectroscopic
metallicity of our work with the previously published ones, extending
the cluster sample out to a projected radius of 117 kpc from the
galaxy's center. We found the  
metallicity gradient for all the confirmed clusters exists with a
slope of $-0.028\pm0.001$ dex kpc$^{-1}$. However, the slope turns to
be $-0.018\pm0.001$ dex kpc$^{-1}$ for all the halo clusters, which is
much shallower. If we only consider the outer halo clusters with
$r_{\rm p}>25$ kpc, the slope becomes $-0.010\pm0.002$ dex kpc$^{-1}$
and if one cluster G001 is excluded from the outer halo sample,  the
slope even could be $-0.004\pm0.002$ dex kpc$^{-1}$. Thus we conclude
that metallicity gradient for M31 outer halo clusters is not
significant, which agrees well with the previous findings. This
  result may imply that for M31 galaxy formation, the ``rapid
  collapsing'' scenario is supported inside the inner halo while the
  ``fragments merging'' scenario is proved in the outer halo of the
  galaxy beyond 25 kpc from the center. It seems that the combination
  of the two scenarios could best explain the galaxy
  formation. However, we still need more observations and further
  study to figure it out.

Besides, it is interesting to note that the halo of M31 galaxy might be
  divided into two parts (by combining the Huxor data): inner halo and
  outer halo from our study. The nature of the two parts of halo
  seem to be different in terms of metallicity gradients of the star clusters,
  which may be due to the different formation mechanisms of the two
  parts. Just like the Milky Way halo from the SDSS/SEGUE data, the Milky Way halo
  could be divided into two parts with different metallicity
  properties based on the observations of a large sample halo
  stars. Therefore, it seems that M31 galaxy and our Galaxy 
  have more similarities than we expected. However,  more
  observational data is required for the further study in the future.

\normalem
\begin{acknowledgements}
  We would like to thank Richard de Grijs for useful discussions. The
  authors are also grateful to the anonymous referee for the helpful
  suggestions and the kind staff at the Xinglong 2.16m 
  telescope for the support during the observations.
  This research was supported by National Natural Science
  Foundation of China through grants 10873016, 10803007, 10633020,
  11003021, 11073027 and 11073032, as well as by the National
  Basic Research Program of China (973 Program) under grant
  2007CB815403. ZF acknowledges a Young Researcher Grant of the
  National Astronomical Observatories, Chinese Academy of Sciences.
\end{acknowledgements}

\appendix                  %%appendicial material is supported

\label{lastpage}


\begin{thebibliography}{99}
\small \setlength{\itemindent}{-3mm} \setlength{\itemsep}{-0.5mm}
\setlength{\baselineskip}{4.5mm}

%% you can type \apj for ApJ, \aap for A&A, \apss for Ap&SS, etc. Please consult
%% the macro raa.cls. You can also find them in aasguide.tex (AASTeX for ApJ, AJ, PASP)
%% Please follow the formats of RAA's references list as demonstrated below:

\bibitem[Ashman \& Bird(1993)]{ash93}Ashman, K.~M., Bird, C.~M.\ 1993,
  \aj, 106, 2281.

\bibitem[Barmby et al.(2000)]{bh00}Barmby, P., Huchra, J., Brodie,
  J., Forbes, D., Schroder, L., Grillmair, C.\ 2000, \aj, 119, 727

\bibitem[Barmby \& Huchra(2001)]{bh01}Barmby, P., \& Huchra, J.~P.\
  2001, \aj, 122, 2458

\bibitem[Barmby et al.(2002)]{bhh02}Barmby, P., Holland, S.,
  Huchra, J.~P.\ 2002, \aj, 123, 1937

\bibitem[Bono et al.(1997)]{bo97}Bono, G., Caputo, F., Cassisi, S.,
  Castellani, V., Marconi, M.\ 1997, \apj, 489, 822

\bibitem[Caldwell et al.(2009)]{cw09} Caldwell, N., Harding, P.,
Morrison, H., Rose, J. A., Schiavon, R., \& Kriessler, J.\ 2009,
\aj,  137, 94

\bibitem[Caldwell et al.(2011)]{cw11}Caldwell, N., Schiavon, R.,
  Morrison, H., Rose, J., Harding, P.\ 2011, \aj, 141, 61

\bibitem[Cardiel et al.(1998)]{car98}Cardiel, N., Gorgas, J., Cenarro, J.
  Gonzalez, J.~J.\ 1998,  \aaps, 127, 597

\bibitem[Carignan et al.(2006)]{cari06}Carignan, C., Chemin, L., Huchtmeier,
  W.~K., Lockman, F.~J.\ 2006, \apj, 641L, 109

\bibitem[Cassisi et al.(1997)]{ccc97}Cassisi S., Castellani M.,
  Castellani, V.\ 1997,  \aap, 317, 10

\bibitem[Djorgovski et al.(1997)]{dj97}Djorgovski S.~G., Gal R.~R.,
  McCarthy J.~K., Cohen J.~G., de Carvalho R.~R., Meylan G.,
  Bendinelli O., Parmeggiani G.\ 1997, \apj, 474, L19

\bibitem[Eggen et al.(1962)]{eggen62}Eggen, O.~J., Lynden-Bell, D.,
  Sandage, A.~R.\ 1962, \apj, 136, 748

\bibitem[Fan et al.(2008)]{fan08}Fan, Z., Ma, J., de Grijs, R.,
  Zhou, X.\ 2008, \mnras, 385, 1973

\bibitem[Fusi Pecci et al.(1994)]{ffp94}Fusi Pecci, F., et al.\ 1994,
  \aap, 284, 349

\bibitem[Galleti et al.(2004)]{gall04}Galleti, S., Federici, L.,
  Bellazzini, M., Fusi Pecci, F., Macrina, S.\ 2004, \aap, 426, 917

\bibitem[Galleti et al.(2005)]{gall05}Galleti, S., Bellazzini, M.,
  Federici, L., Fusi Pecci, F.\ 2005, \aap, 436, 535

\bibitem[Galleti et al.(2006)]{gall06} Galleti, S., Federici, L.,
  Bellazzini, M., Buzzoni, A., Fusi Pecci, F.\ 2006, \aap, 456, 985

\bibitem[Galleti et al.(2007)]{gall07} Galleti, S., Bellazzini, M.,
  Federici, L., Buzzoni, A., Fusi Pecci, F.\ 2007, \aap, 471, 127

\bibitem[Galleti et al.(2009)]{gall09}Galleti, S.,
  Bellazzini, M., Buzzoni, L., Federici, L., Fusi Pecci, F.\ 2009,
  \aap, 508, 1285

\bibitem[Hammer et al.(2007)]{ham07}Hammer, F., Puech, M., Chemin, L.,
  Flores, H., Lehnert, M. D.\ 2007, \apj, 662, 322	 

\bibitem[Hou et al.(2009)]{hou09}Hou, J., Yin, J., Boissier, S.,
  Prantzos, N., Chang, R.~X., Chen, L.\ 2009, IAU Symposium, 254, 27	 

\bibitem[Huchra et al.(1982)]{hsv82}Huchra, J.~P., Stauffer, J., van
  Speybroeck, L.\ 1982, \apj, 259, L57

\bibitem[Huchra et al.(1991)]{hbk91}Huchra J.~P., Brodie J. P., Kent
  S. M.\ 1991, \apj, 370, 495

\bibitem[Huxor et al.(2004)]{h04}Huxor, A., Tanvir, N.~R., Irwin, M.~J.,
  Ferguson, A.~M.~N., Ibata, R.~A., Lewis, G.~F., Bridges, T.\ 2004,
  in ASP Conf. Ser. 327, Satellites and Tidal Streams, ed. F. Prada,
  D. Martinez-Delgado, \& T. Mahoney (San Francisco: ASP), 118

\bibitem[Huxor et al.(2005)]{h05}Huxor, A.~P., Tanvir, N.~R., Irwin, M.~J.,
  Ibata, R., Collett, J.~L., Ferguson, A.~M.~N., Bridges, T., Lewis,
  G.~F.\ 2005,  \mnras, 360, 1007

\bibitem[Huxor(2007)]{h07}Huxor, A.\ 2007, Ph.D. Thesis, Univ. of
  Hertfordshire, UK

\bibitem[Huxor et al.(2011)]{h11}Huxor, A.~P., Ferguson, A.~M.~N.,
  Tanvir, N.~R., Irwin, M.~J., Mackey, A.~D., Ibata, R.~A., Bridges,
  T., Chapman, S.~C., Lewis, G.~F.\ 2011, arXiv1102.0403H

\bibitem[Jester et al.(2005)]{jes05} Jester, S., et al.\ 2005, \aj,
  130, 873

\bibitem[Ma et al.(2010)]{ma10}Ma, J., Wu, Z., Wang, S., Fan, Z.,
  Zhou, X., Wu, J., Jiang, Z., Chen, J.\ 2010, \pasp, 122, 1164

\bibitem[Macri(2001)]{mac01}Macri, L.~M.\ 2001, \apj, 549, 721

\bibitem[Mackey et al.(2006)]{mac06}Mackey, A.~D., et al.\ 2006,
  \apj, 653, L105

\bibitem[Mackey et al.(2007)]{mac07}Mackey, A.~D., et al.\ 2007,
  \apj, 655, L85

\bibitem[Mackey et al.(2010)]{mac10}Mackey, A.~D., et al.\ 2010,
  \mnras, 401, 533
  
\bibitem[Maraston(1998)]{mar98}Maraston, C.\ 1998, \mnras, 300, 872

\bibitem[Massey et al.(1988)]{mass88}Massey, P., Strobel, K., Barnes,
  J. V., Anderson, E.\ 1988, \apj, 328, 315

\bibitem[McConnachie et al.(2005)]{mcc05}McConnachie, A.~W., Irwin,
  M.~J., Ferguson, A.~M.~N., Ibata, R.~A., Lewis, G.~F., Tanvir, N.\
  2005, \mnras, 356, 979

\bibitem[Peacock et al.(2010)]{pc10} Peacock, M.~B., Maccarone, T.~J.,
  Knigge, C., Kundu, A., Waters, C.~Z., Zepf, S.~E., \& Zurek, D.~R.\
  2010, \mnras, 402, 803

\bibitem[Peng et al.(2006)]{peng06}Peng, E., et al.\ 2006,
  \apj, 639, 838

\bibitem[Perina et al.(2010)]{per10}Perina, S., et al.\ 2010,
  \aap, 511, 23

\bibitem[Perrett et al.(2002)]{per02}Perrett, K.~M., Bridges, T.~J.,
  Hanes, D.~A., Irwin, M.~J., Brodie, J.~P., Carter, D., Huchra, J.~P.,
  Watson, F.~G.\ 2002, \aj ,123, 2490

\bibitem[Racine(1991)]{rac91}Racine, R.\ 1991, \aj, 101, 865

\bibitem[Rich et al. (2005)]{rich05}Rich, R.~M., Corsi, C.~E., Cacciari,
  C., Federici, L., Fusi Pecci, F., Djorgovski, S.~G., Freedman, W.~L.\
  2005, \aj, 129, 2670

\bibitem[Salpeter(1955)]{sal}Salpeter, E.~E.\ 1955, \apj, 121, 161

\bibitem[Searle \& Zinn(1978)]{sz78}Searle L., Zinn R.\ 1978, \apj,
225, 357

\bibitem[Stanek \& Garnavich(1998)]{sg98}Stanek, K.~Z., Garnavich,
  P.~M.\ 1998, \apj, 503, 131

\bibitem[Thomas et al.(2003)]{tmb}Thomas, D., Maraston,
  C., Bender, R.\ 2003, \mnras, 339, 897

\bibitem[Thomas et al.(2010)]{tmj}Thomas, D., Maraston,
  C., Johansson, J.\ 2010, arXiv1010.4569.

\bibitem[Thomas et al.(2004)]{tmk}Thomas, D., Maraston,
  C., Korn, A.\ 2004, \mnras, 351, L19

\bibitem[van den Bergh(1969)]{van69}van den Bergh, S.\ 1969,
\apjs, 19, 145

\bibitem[Worthey et al.(1994)]{w94}Worthey, Guy, Faber, S.~M.,
  Gonzalez, J.~J., Burstein, D.\ 1994, \apjs, 94, 687

\bibitem[Worthey \&  Ottaviani(1997)]{wo97}Worthey, Guy, Ottaviani,
  D.~L.\ 1997, \apjs, 111, 377

\bibitem[Yin et al.(2009)]{yin09}Yin, J., Hou, J.~L., Prantzos, N.,
  Boissier, S., Chang, R.~X., Shen, S.~Y., Zhang, B.\ 2009, \aap, 505,
  497   

\end{thebibliography}
\end{document}